%% file: main.tex
\documentclass[conference]{IEEEtran}
\IEEEoverridecommandlockouts
\usepackage{cite}
\usepackage{amsmath,amssymb,amsfonts}
\usepackage{algorithmic}
\usepackage{graphicx}
\usepackage{textcomp}
\usepackage{adjustbox}
\usepackage{makecell}
\usepackage{url}

\usepackage{comment}
\usepackage{xspace}
\usepackage[english]{babel}
\usepackage{multicol}
\usepackage[table,xcdraw]{xcolor}
\def\BibTeX{{\rm B\kern-.05em{\sc i\kern-.025em b}\kern-.08em
    T\kern-.1667em\lower.7ex\hbox{E}\kern-.125emX}}
\begin{document}

\title{Clinical Landscape of COVID-19 Testing: Difficult Choices}

\author{\IEEEauthorblockN{Darshan Gandhi\IEEEauthorrefmark{1},
                          Sanskruti Landage\IEEEauthorrefmark{1},
                          Joseph Bae\IEEEauthorrefmark{1}\IEEEauthorrefmark{2},
                          Sheshank Shankar\IEEEauthorrefmark{1},
                          Rohan Sukumaran\IEEEauthorrefmark{1},
                          Parth Patwa\IEEEauthorrefmark{1},\\
                          Sethuraman T V\IEEEauthorrefmark{1},
                          Priyanshi Katiyar\IEEEauthorrefmark{1},
                          Shailesh Advani\IEEEauthorrefmark{1},
                          Rohan Iyer\IEEEauthorrefmark{1},
                          Sunaina Anand\IEEEauthorrefmark{1},
                          Aryan Mahindra\IEEEauthorrefmark{1},
                          Rachel Barbar\IEEEauthorrefmark{3},\\
                          Abhishek Singh\IEEEauthorrefmark{3},
                          Ramesh Raskar\IEEEauthorrefmark{1}\IEEEauthorrefmark{3}
                          }\\
        \IEEEauthorblockA{\IEEEauthorrefmark{1}\textit{PathCheck Foundation},\IEEEauthorrefmark{3}\textit{MIT Media Lab}\\ Cambridge, MA, USA\\ \IEEEauthorrefmark{2}\textit{Renaissance School of Medicine}\\ Stony Brook, NY, USA\\}
        Corresponding Author: darshan.gandhi@pathcheck.org}

\maketitle

\begin{abstract}
The coronavirus disease 2019 (COVID-19) pandemic has spread rapidly across the world, leading to enormous amounts of human death and economic loss. Until definitive preventive or curative measures are developed, policies regarding testing, contact tracing, and quarantine remain the best public health tools for curbing viral spread. Testing is a crucial component of these efforts, enabling the identification and isolation of infected individuals. Differences in testing methodologies, time frames, and outcomes can have an impact on their overall efficiency, usability and efficacy. In this early draft, we draw a comparison between the various types of diagnostic tests including PCR, antigen, and home tests in relation to their relative advantages, disadvantages, and use cases. We also look into alternative and unconventional methods. Further, we analyze the short-term and long-term impacts of the virus and its testing on various verticals such as business, government laws, policies, and healthcare.  
\end{abstract}


\input{content/introduction}
\input{content/related_work}
\input{content/native_methodology}
\input{content/alternative_testing_methods}
\input{content/unconventional_testing_methodologies}

\input{content/effects}
\input{content/future_possibilities}
\input{content/conclusion}

\section*{Acknowledgements}
We are grateful to Riyanka Roy Choudhury, CodeX Fellow, Stanford University, Adam Berrey, CEO of PathCheck Foundation, Dr. Brooke Struck, Research Director at The Decision Lab, Canada and Vinay Gidwaney, Entrepreneur and Advisor, PathCheck Foundation for their assistance in discussions, support and guidance in writing of this paper.

\bibliography{refs}\nocite{*}
\bibliographystyle{plain}

\end{document}

%% file: content/introduction.tex
\section{Introduction}
Severe acute respiratory syndrome coronavirus 2 (SARS-CoV-2) was identified as a novel type of coronavirus after the outbreak in the Wuhan region of China in late 2019. On March 11, 2020, the World Health Organization (WHO) declared COVID-19, the disease caused by this RNA virus, as a pandemic. COVID-19 is a disease with a relatively high transmission rate, and is spread primarily by person-to-person contact through respiratory droplets. As of November 5th 2020, COVID-19 has led to more than 48 million cases and 1 million deaths globally. Diagnosis of COVID-19 among patients involves testing through laboratory testing, clinical examination and diagnostics (X-ray, Chest-CT and others). Early positive test identification enables isolation of infected patients in order to reduce transmission \cite{30}. Notably, a large group of individuals identified as infected with SARS-CoV-2 are asymptomatic. This suggests the necessity for frequent testing of all individuals in a population, which in turn requires more cost efficient and rapid testing platforms .

Presently, reverse transcription polymerase chain reaction (RT-PCR) assays are the most widely used method of COVID-19 detection. RT-PCR identifies the existence of viral RNA in biological specimens from patients \cite{36}. In RT-PCR testing the most time-consuming steps are purification of RNA and reverse transcription. Because it detects viral RNA, RT-PCR is extremely sensitive and specific for SARS-CoV-2 infection detection. Rapid antigen tests are another common form of testing in the setting of COVID-19, and detect viral proteins in patient samples. These tests are much cheaper, faster, and more convenient than RT-PCR tests, but are generally less sensitive. Although less sensitive, antigen tests can be used to identify the most infected patients in the affected region to control the spread of COVID-19 \cite{38}.

In this early draft, our goal is to address the current scenario of  COVID-19 testing methodologies and summarize the implementation, cost and production estimates, and finally effects on different sectors of society of these tests. Relevant related work in surveys, comparative studies and methods to mitigate the COVID-19 crisis is discussed in the Related Work section. We then move to an in-depth discussion of the various current testing methods stratified by molecular, serological and alternative/unconventional approaches in the Native Methodologies in Clinical Testing section. We further dive deeper to understand the implications of testing and the impacts it has on privacy, communication, the economy, etc. in the Effect of Testing section.

\begin{figure}[!ht]
    \includegraphics[width=1\linewidth]{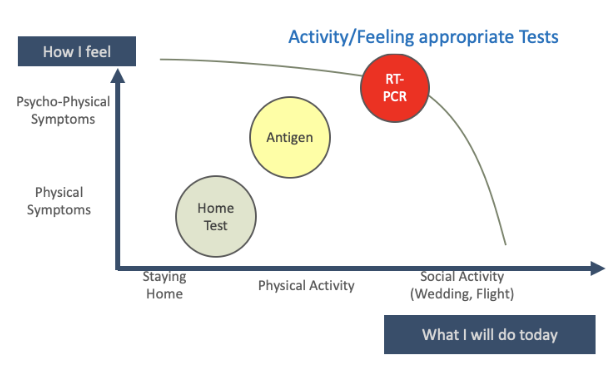}
    \centering
    \caption{Symptom profiles and patient behaviours appropriate for various COVID-19 testing methodologies. 
    }
    \label{fig:activity_feeling_guide}
\end{figure}

%% file: content/related_work.tex
\section{Related Work}

\begin{figure}
    \includegraphics[width=1\linewidth]{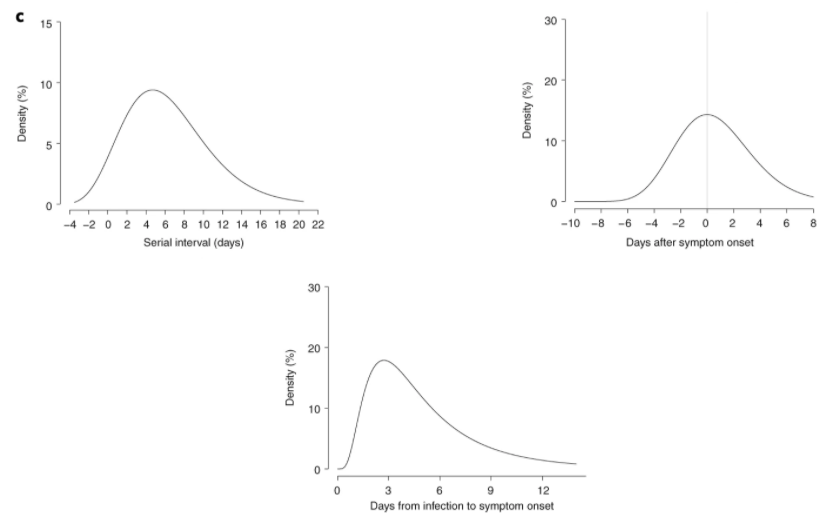}
    \centering
    \caption{Estimated serial interval distribution (top left), inferred infectiousness profile (top right) and assumed incubation period (bottom middle) of COVID-19 (waiting for permission) \cite{44}}
    \label{fig:serial_interval_distribution}
\end{figure}

\begin{figure}
    \includegraphics[width=1\linewidth]{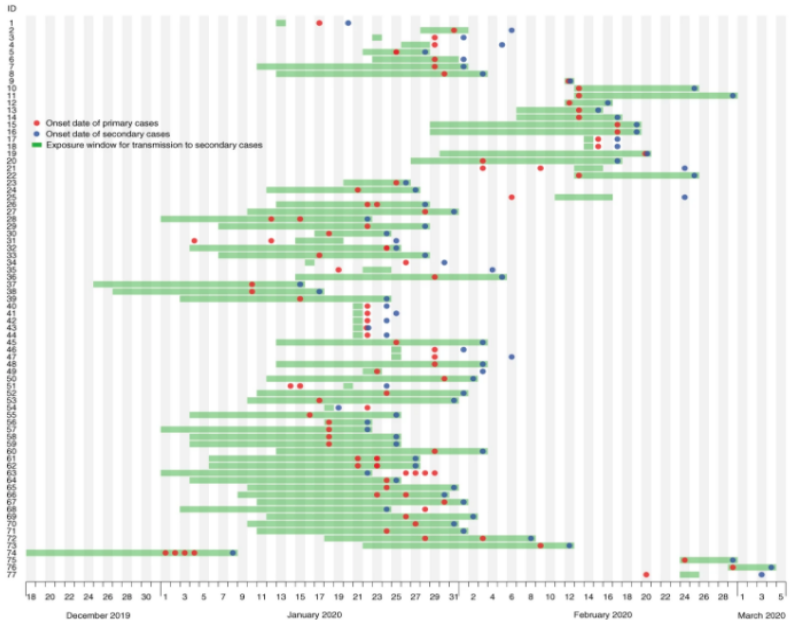}
    \centering
    \caption{Human to Human Transmission: Temporal dynamics in viral shedding and transmission of COVID-19 (waiting for permission) \cite{44}}
    \label{fig:human_to_human_transmission}
\end{figure}

Fig 1 elaborates on temporal patterns of viral shedding in 94 patients with laboratory-confirmed COVID-19 and modeled COVID-19 infectiousness profiles from a separate sample of 77 infector–infectee transmission pairs. They observed the highest viral load in throat swabs at the time of symptom onset, and inferred that infectiousness peaked on or before symptom onset. The authors have estimated that 44\% (95\% confidence interval, 30–57\%) of secondary cases were infected during the index cases’ presymptomatic stage, in settings with substantial household clustering, active case finding and quarantine outside the home. Disease control measures should be adjusted to account for probable substantial presymptomatic transmission \cite{44}.

Tang Wi et al. discusses the current landscape of COVID-19 in different windows, namely pre-analytical, analytical, and post-analytical. The pre-analytical window discusses the importance of collecting the right specimen at the right time in COVID-19 infection. The analytical window seeks to weigh the relative advantages of different testing approaches including RT-PCR and antigen tests. Finally the post-analytical landscape discusses how tests might be understood with relevant molecular or serological  bottlenecks. Our paper aims to go further in providing a comprehensive analysis of both the various tests available as well as their relative advantages in different use cases. We also discuss various secondary and alternative testing methods in addition to serological approaches \cite{16}.

Mina et al. examines the current  testing scenario to explain the flaws in the current testing landscape. They primarily claim that the more accurate RT-PCR tests used may be primarily detecting non-infectious individuals with low viral loads. Mina et al. also suggests that a low sensitivity test done at more frequent intervals might be more efficient in understanding viral spread than limited, high sensitivity tests might be able to do \cite{1}.

Marac et al. does a comprehensive review on molecular and serological tests and also talks about the need to upscale in-vitro diagnostic assays.  The paper also shows that these different tests could help inform healthcare providers and policy makers, especially in understanding the efficacy of tests and in making efficient policy decisions \cite{32}.

Ngyun, T Thanha discusses the AI methods that are made to understand and fight against the ongoing pandemic. The paper is a survey of methods in image analysis, text understanding and information extraction, data analytics, IoT, etc.. An overview of the relevant data sources and a description of 13 sub groups of problems where AI and relevant methods can potentially be useful are also presented \cite{18}.

%% file: content/native_methodology.tex
\section{Native Methodologies of Clinical Testing}

\subsection{Need for Testing}

Given that many people infected with COVID-19 continue to be asymptomatic and potential carriers of the virus, testing plays a crucial role in identifying these individuals to prevent further spread through contact tracing and isolation. Testing can be carried out to identify infected cases, most importantly helping to identify individuals who are infected and isolate them. With the availability of numerous testing practices, there are many methods with different analytical performances in terms of sensitivity and specificity. Because viral load is often correlated with the probability of a test, patients with an early infection may often not be detected. After 3-4 days following infection, there is an exponential increase in viral load, generally seeming to result in symptom development. These can range from mild infection to adverse and serious events involving lung disease, acute respiratory distress syndrome (ARDS), and other non-pulmonary manifestations.  

\subsection{Diagnostic Tests}
Testing for SARS-CoV-2 can be largely classified into two broad categories: diagnostic testing attempts to identify viral presence, whereas antibody/serological tests identify the immune response generated to the virus \cite{10}.

The gold standard method for diagnostic testing involves the use of reverse transcription-polymerase chain reaction (RT-PCR). RT-PCR tests identify viral genomic material from human samples typically taken from the nasopharyngeal cavity, throat, saliva, and so on.  Specimens obtained from patients must therefore be sent to centralized laboratories for storage and testing \cite{14,15}. The results of these tests can be obtained in a few hours after specimens arrive at the laboratory, but the total turnaround time to patients is often between 3-7 days. The test is highly sensitive and specific in nature and is generally the gold standard at the moment.  Reliable methods of evaluating the objective accuracy, sensitivity, and specificity of RT-PCR tests have not been developed, but it is assumed that in ideal laboratory conditions these tests are able to detect SARS-CoV-2 genetic material with high degrees of sensitivity and specificity \cite{17, 26, 29, 31}.

\begin{figure}[!ht]
    \includegraphics[width=1\linewidth]{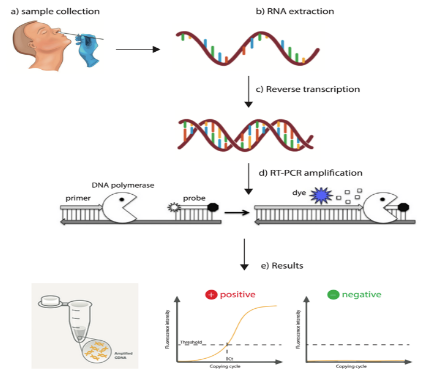}
    \centering
    \caption{Steps in the RT-PCR test: a) A patient specimen is obtained (commonly through nasopharyngeal swab) b) viral RNA is retrieved and c) reverse transcription is used to create complimentary DNA (cDNA). d) DNA primers are then affixed to cDNA and are used for amplification. Fluorescence is created following degradation of specialized DNA probes which increase as copies of the viral cDNA are made. e) Once the detected fluorescence level crosses a threshold, a positive test is identified (waiting for permission) \cite{39}.}
    \label{fig:rt_pcr_diagram}
\end{figure}

Rapid Antigen Detection Tests (RADT) is a rapid way of testing to detect the presence or absence of SARS-CoV-2 in a point-of-care setting. These tests identify the presence of virus-specific proteins (antigens) in collected specimens from patients. Specimens are generally collected from the same regions as those used for RT-PCR (nasopharyngeal, nasal, throat, etc), but testing can be done in a rapid manner and is often done on-site rather than at specialized laboratories [25]. Specimens are analyzed via immunological assays for viral particles and can be done in many non-specialized laboratory settings. Results for RADT can be obtained as quickly as 30-60 minutes, but the sensitivity of these tests is generally lower than RT-PCR (84-97\% when compared with RT-PCR results). However, these tests are considered very specific (90-98\% when compared with RT-PCR). Because of this, a patient testing positive by RADT will generally be considered to be infected with the virus whereas a patient testing negative may be referred to an RT-PCR test if there is suspicion of infection. Due to their speed and cost-effectiveness, RADT can be used in point-of-care settings (doctors, hospitals) and can also be used as a public health intervention for rapid detection and estimation of disease burden in a community setting.

\begin{figure}[!ht]
    \includegraphics[width=1\linewidth]{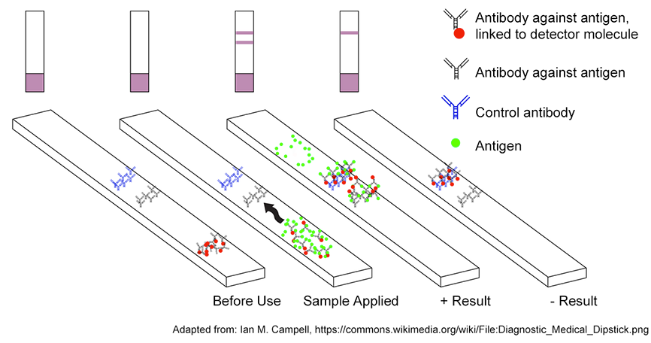}
    \centering
    \caption{Diagram briefly describing the function of medical diagnostic dipsticks (waiting for permission) \cite{40}.}
    \label{fig:radt_diagram}
\end{figure}

Lateral flow assay (LFA) antigen tests are another example of rapid point-of-care testing. The test analyzes plasma, blood, or other patient samples and is capable of detecting viral antigen. These tests do not require laboratory analysis and can be performed in any clinical setting. These tests have relatively low sensitivities but are both fast and cost-effective. Due to these reasons, these sometimes serve as an alternative and a useful tool in identifying the prevalence of COVID-19 spread in a community \cite{14}.

\begin{figure}[!ht]
    \includegraphics[width=1\linewidth]{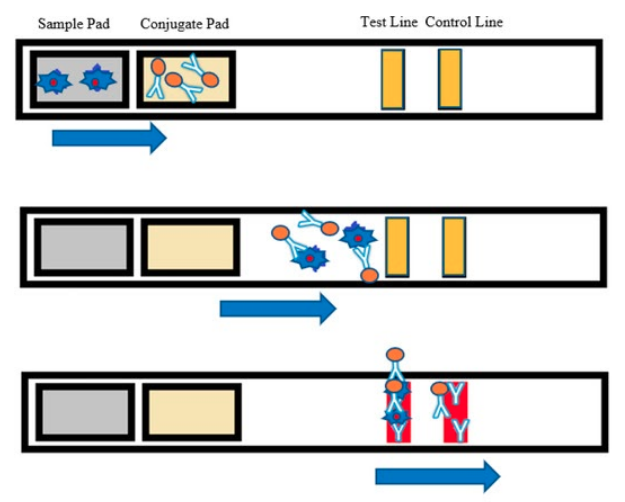}
    \centering
    \caption{LFA Strip architecture where the analyte is detected in the test line, and red control line indicates the test was performed (waiting for permission)  \cite{41}.}
    \label{fig:lfa_strip_diagram}
\end{figure}

The RT-PCR test employs the amplification of small amounts of viral RNA particles in patient specimens. \textbf{Cycle Thresholding} (CT)  \cite{23} is the concept of amplifying the virus ‘x’ number of times, where ‘x’ can be any number in order to understand the viral concentration (viral load). Lower CT values mean that fewer amplification cycles were required to detect viral RNA, indicating that the patient sample had a high viral load. Alternatively, high CT Values indicate that the viral load is comparatively low and hence additional amplifications are required for detection.There is no definite threshold for clearing the individual for the risk of spreading the virus. Some research has indicated that CT values can be correlated to infectivity, with higher values indicating a lower chance that a person might transmit COVID-19. CT values also differ based on parameters such as age, gender, the prevalence of comorbidities, or concurrent infection \cite{24}.

\begin{figure}[!ht]
    \includegraphics[width=1\linewidth]{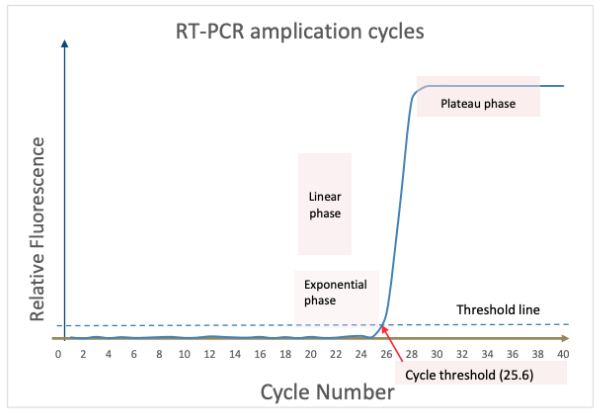}
    \centering
    \caption{Stages for RT-PCR Post Run Analysis (waiting for permission) \cite{23}.}
    \label{fig:lfa1_strip_diagram}
\end{figure}

\textbf{Rethinking the Test Sensitivity}: When considering the relative advantages and disadvantages of the various diagnostic COVID-19 tests approved by the FDA for emergency use, it is important to understand the intended use case of the test. Mina et al. \cite{1}, suggests that low sensitivity, low price tests such as LFA antigen tests might most effectively be used in public health scenarios in which the desired outcome is outbreak suppression. By testing each individual in a community daily with these tests, Mina et al., argues that patients with high viral loads (and therefore potentially higher infectivity) will be immediately recognized by low sensitivity tests and can then be quarantined to prevent disease spread. While this does seem to be an attractive testing policy for reducing outbreaks, it relies on an increased understanding into the relationship between viral load, infectivity, and symptom manifestation in the setting of COVID-19. Further, it is likely that this testing paradigm would still need to be supplemented by higher sensitivity testing in vulnerable populations where even low viral load COVID-19 infection could result in severe outcomes.

\begin{figure}[!ht]
    \includegraphics[width=1\linewidth]{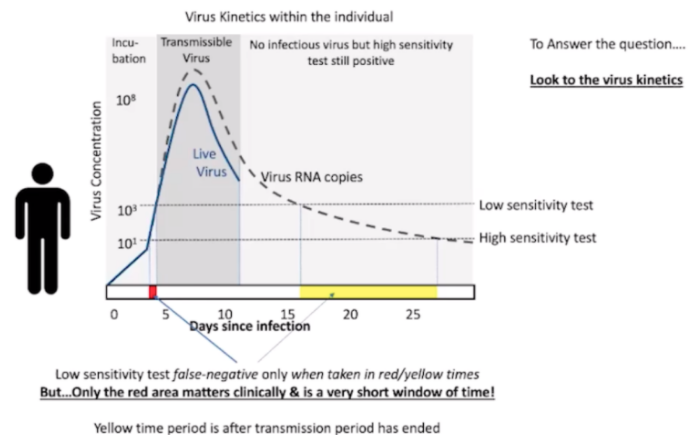}
    \centering
    \caption{Virus Kinetics within the Individual (waiting for permission) \cite{1}}
    \label{fig:virus_kinetics}
\end{figure}

\subsection{Antibody Tests}
Antibody tests are done after recovery from COVID-19 to check whether a patient has developed immunity to COVID-19 through measurements of IgM or IgG antibodies against the virus. Antibodies are a physiological immune response against SARS-CoV-2 infection, and their presence might indicate that an individual has recently been infected with the virus or has developed long term immunity to the same. The information provided by antibody testing is highly variable from patient to patient and depends significantly on when the test was acquired relative to infection with SARS-CoV-2. Antibody responses can take up to weeks to be detectable following viral infection, and some patients might not have detectable levels despite a previous infection. Alternatively, there is evidence that positive test results can be observed for patients without recent SARS-CoV-2 infection due to the presence of antibodies to similar viruses. These tests are not meant to be diagnostic in nature. Instead, antibody testing is often done to determine whether an individual had previously been infected by SARS-CoV-2. This can be useful in the setting of determining eligible plasma donors for COVID-19 treatments, or in other similar scenarios  \cite{11, 17, 34}.

\begin{figure*}[t!]
\resizebox{\textwidth}{!}{
    \includegraphics[]{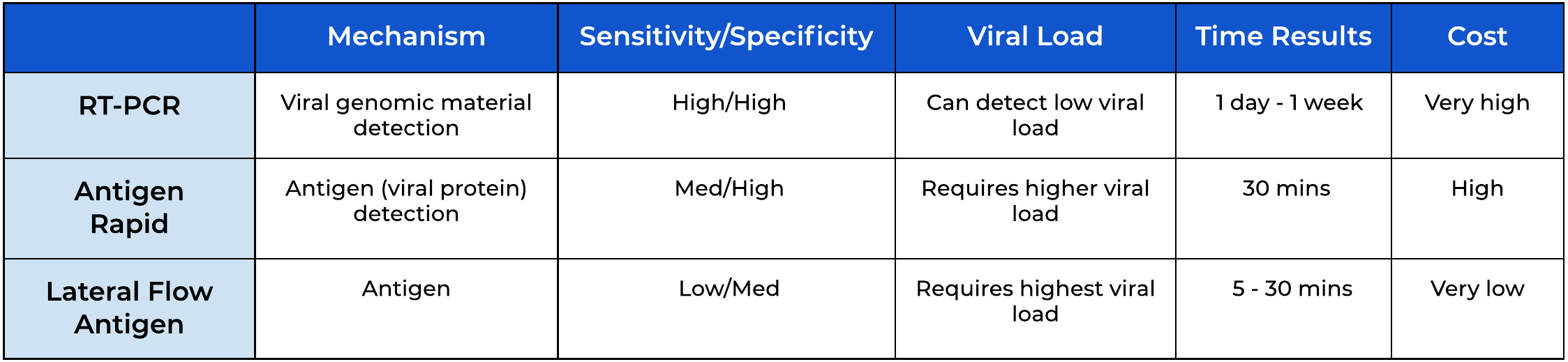}
    }
    \centering
    \caption{Comparison between Native Testing Methodologies}
    \label{fig:comparison_table}
\end{figure*}

\begin{figure}[!ht]
    \includegraphics[width = 1\linewidth]{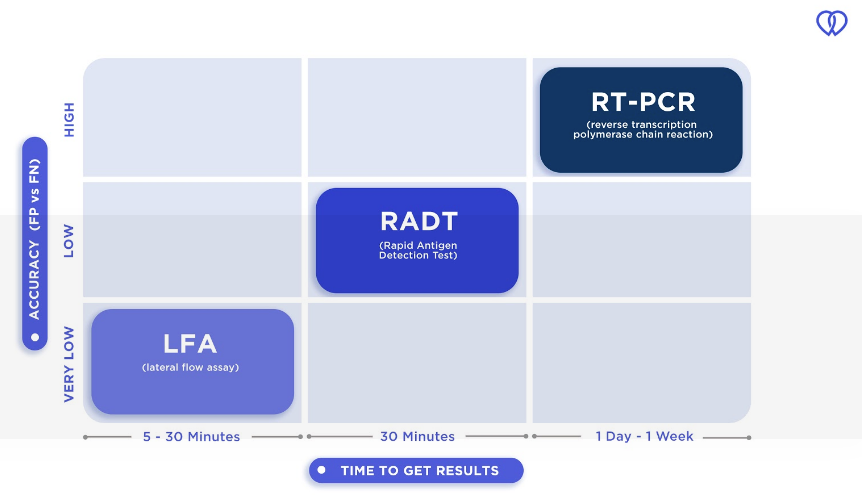}
    \centering
    \caption{Comparative Analysis Between Native Testing Methodologies}
    \label{fig:my_label}
    
\end{figure}

\subsection{Current Scenario of Testing}
Testing guidelines have varied over the course of the COVID-19 pandemic, generally dependent on the availability of tests. Initially, testing supply shortages in the United States necessitated strict rationing of testing kits to only those who are sick and mandating that individuals be recommended by a physician in order to receive a test. Currently, the CDC recommends that individuals with symptoms of infection, recent contact with someone with COVID-19, or a recommendation from a healthcare provider be tested for infection with SARS-CoV-2. Additionally, workplaces require essential workers and other non-essential workers to have repeat COVID-19 tests before arriving at their workplace. Molecular tests including RT-PCR are recommended for diagnosis of COVID-19, as these are known to provide the most specific and sensitive results. 

Errors in testing can have dramatic effects. False-negative results might delay the quarantine of an individual, increasing the likelihood of disease spread. False-positive results might cause undue stress for a patient and cause loss of access to economic, social, and personal support in lieu of quarantine. It is important to note that comparative analysis between the various testing methodologies with respect to their accuracy, false negative rate and costing are not completely dependable since there are many players in the market who manufacture the testing kits to be used for the process and also the inherent disparity that can be observed in the laboratory vs clinical performance of a test. 

Home test kits are a recent development and are becoming increasingly prevalent. The majority of home tests collect samples of either saliva (in a collection tube) or nasal swabs (through cotton swabs). Using these kits, an individual can ship their samples to testing labs directly from their house. After the sample is received by the lab, results are typically delivered within 48-72 hours  \cite{12}. Although these tests generally suffer from lower accuracies than RT-PCR, they are attractive due to their convenience and relatively low price point. 

There remains a need to develop strict guidelines for social distancing and mobility including use of PPE and masks when visiting public places.

\subsection{Activity-Feeling Guide}

Getting tested is not a binary decision. There are many decisions that need to be made for uninformed user-specific guidance, such as an Activity-Feeling Matrix, which is a helpful tool. It is important to promote low friction home tests for low-stakes situations and drive behavior to optimal risk-anxiety levels. 

For example, depending on the individual's level of activity (i.e. work from home vs working in the community) level and frequency of tests might vary in terms of intensity and follow up. There can be an observed shift from the less accurate tests such as home tests to the far more accurate ones like the RT-PCR if an individual wishes to attend social functions or has a need for travel.

\subsection{Assessment of Risks and Incident Prevention}
Risk management is a highly understood field, and well-established strategies make use of multi-layer interventions in order to prevent undesirable outcomes. 

\begin{figure}[!ht]
    \includegraphics[width=1\linewidth]{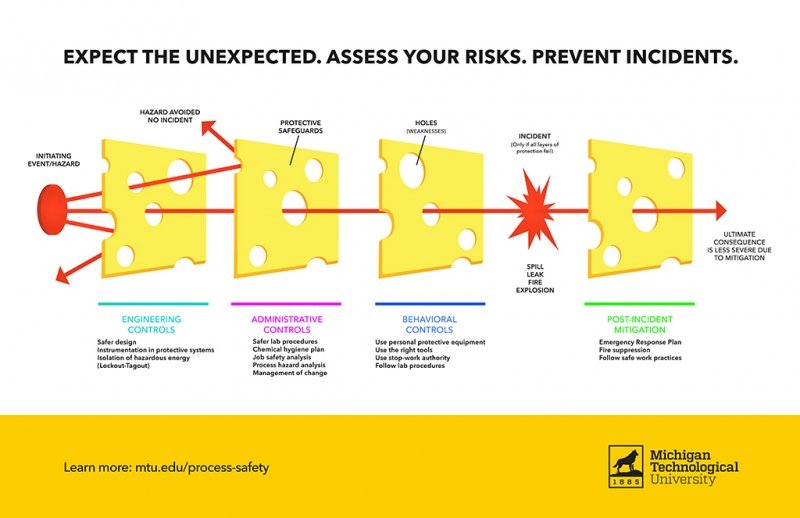}
    \centering
    \caption{Swiss cheese model of process safety (waiting for permission)  \cite{42}}
    \label{fig:risks}
\end{figure}

This same approach can be adopted in the setting of COVID-19, in which multiple interventions can be simultaneously leveraged to reduce disease spread. These strategies might include maintaining social distancing practices, employing effective ventilation systems in buildings, wearing masks, washing and sanitizing hands at regular intervals, having quicker and efficient testing procedures, and improved contact tracing procedures in order to quickly contain outbreaks of the disease. Further, strict follow-up of individuals who are either quarantined due to positive test or exposed to a COVID-19 positive individual remains crucial for successful reduction in the burden of COVID-19.

\begin{figure}[!ht]
    \includegraphics[width=1\linewidth]{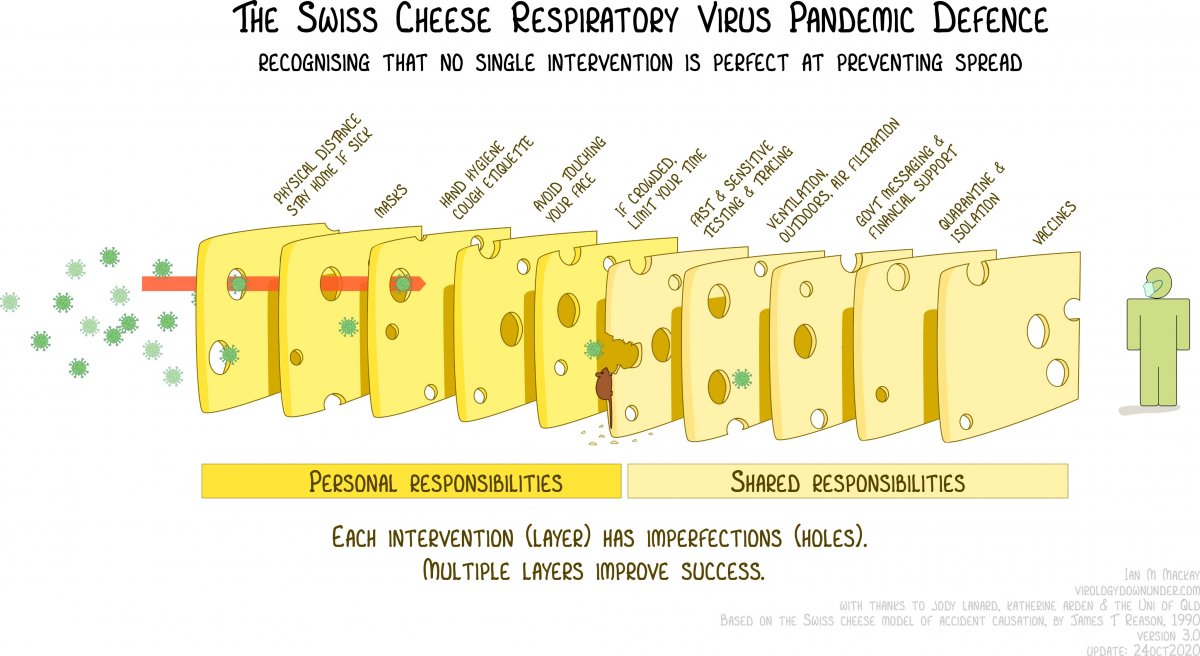}
    \centering
    \caption{Swiss Cheese Respiratory Virus Pandemic Defense (Swiss Cheese Respiratory Virus Pandemic Defense) (waiting for permission) \cite{43}}
    \label{fig:swiss_cheese}
\end{figure}

%% file: content/alternative_testing_methods.tex
\section{Alternative Testing Methods}
\subsection{Introduction}
It is integral to expand testing capabilities to manage the further spread of COVID-19. Along with this expansion, there is a need for developing alternative strategies to identify and isolate asymptomatic and presymptomatic individuals. Efficient pipelines for specimen collection and analysis are crucial. 

A variety of factors must be considered when selecting a COVID-19 test including cost, accuracy, availability, dissemination, and speed of testing results. Depending on the use case, each of these factors might be weighted in a different manner. There is an increase in alternative testing methods as emergency use authorizations have been provided by the FDA, enabling multiple community-level testing strategies to control COVID-19.

\subsection{Pool Testing}
In pooled tests, several samples are mixed together in a batch in order to test specimens from multiple individuals simultaneously. This allows for greater testing capabilities without using more resources. In this process there is an increased chance of positive sample dilution, potentially decreasing test sensitivity. According to FDA guidelines, test performance following the pooling of samples should have a positive predictive agreement (PPA) of greater than 85\% when compared with tests run on individual samples \cite{8}.  This type of testing substantially reduces the costs of testing and is a faster method for testing multiple individuals. This enables estimations of COVID-19 prevalence in group settings such as institutions or communities. The ideal use case for these tests is in the setting of relatively sparse COVID-19 positivity/low prevalence \cite{13,15}.

\subsection{Wastewater Testing}
Experiences with other viral diseases have shown that monitoring wastewater enables effective surveillance of COVID-19 by tracing pathogen levels in sewage. This can effectively show the presence of pathogens in an entire community, providing a sensitive signal to detect whether the transmission is increasing or declining.

Wastewater testing could have many benefits with regards to the detection of COVID-19 as it is cost-effective to survey the dynamics of communities. Other forms of epidemiological indicator biases are effectively avoided and it enables the collection of data even from people who don’t have access to healthcare. 

After excretion in feces, the virus collects in municipal wastewater channels. Monitoring of this wastewater at treatment plants can be achieved by detecting viral RNA. 

%% file: content/unconventional_testing_methodologies.tex
\section{Unconventional Testing Methodologies}

\subsection{COVID-19 and Animals}
In most research works related to healthcare and medicine animals are very essential segment of the cycle, especially in zoonotic viruses such as SARS-CoV-2. However, thorough testing on animals has not been performed yet; only a few species (such as ferrets, dogs, birds, cats, hamsters and minks) have shown to be susceptible to the virus. Preliminary research has shown high interspecies transmission rates in ferrets, cats, and golden hamsters, but significantly lower rates in dogs. Furthermore, animals such as ducks, chickens, mice and pigs do not seem to be able to contract or transmit the virus\cite{35}. Further experiments can be performed in these species to understand the mechanisms by which other animals might recover from infection, potentially informing treatment development for humans\cite{22}.

\subsection{Reverse transcription loop-mediated isothermal amplification (RT-LAMP)}
Researchers from the Journal of Science Translational Medicine proposed the use of a technique called reverse transcription loop-mediated isothermal amplification (RT-LAMP)\cite{2} carried out at a constant temperature with a different set of reagents than those used in RT-PCR. Here, simple equipment like plate scanners with spectrophotometric quantification, mobile phone cameras, copy machines, or office scanners can be used for immunofluorescent detection. Expensive special equipment such as thermal cyclers with real-time fluorescence measurements is not required for an RT-LAMP assay to perform, because within 30 minutes after the start of incubation (at 65°C), positive samples can be determined by a color change from red to yellow\cite{14,17}.

This testing is less sensitive than a quantitative RT-PCR, however, it could be beneficial in the testing of large groups of people because of its potential to be a simple, scalable, and broadly applicable testing method.

\subsection{SalivaDirect}
SalivaDirect is a testing method that relies on a similar molecular process as RT-PCR. Compared to RT-PCR, this method is less expensive and places less stress on the supply chain. In addition, RT-PCR requires a trained person to collect nasopharyngeal (NP) swabs, not only adding a logistical barrier, but also putting the performer at risk of getting infected. Saliva samples are easier to collect and less uncomfortable, encouraging people to get tested frequently. However, saliva tests were found to have lower sensitivity than a nasopharyngeal swab test\cite{37}. 

\subsection{LamPORE}
LamPORE combines barcoded multi-target amplification, real-time nanopore sequencing, and 15-minute barcode library preparation\cite{3}. It enables a larger number of samples to be rapidly tested/screened for the presence or absence of SARS-CoV-2, the virus causing COVID-19. Starting with extracted RNA, results for 12-96 samples can be obtained in less than 2 hours\cite{2}.

This method not only fits both large-scale and small-scale laboratory environments, but can also potentially analyze thousands of samples daily using a single instrument. LamPORE is dependent on up-to-date workflow, including automatic handling of all samples and an amalgamation with the information management systems of the laboratory.

%% file: content/effects.tex
\section{Effect of the Testings}

\subsection{Quarantine}
Generally, an individual takes a COVID-19 test because they are either showing symptoms of the virus, or due to mandated workspace regulations. Until test results are received, the individual must stay isolated. If an individual receives a negative result from a lower sensitivity point-of-care or home test, it is recommended that they seek another negative result via RT-PCR if infection is still suspected. During quarantine, the patient should keep away from others, wear a face covering, follow strict hand-hygiene, and make an effort to utilize different utensils, instruments and tools from other members of his/her household \cite{30}.

In case of an emergent need for the patient to break quarantine, they should maintain at least 6 feet distance from others. In addition, caretakers and family members exposed to an individual with COVID-19 should follow similar guidelines (including testing, quarantine, and preventive techniques).

\subsection{Privacy}
COVID-19 testing practices involve the collection and distribution of a significant amount of personal information. Often an individual’s name, race, location, travel history, and past health records are collected. This dependence on sensitive patient data provides many potential privacy concerns related to COVID-19 testing. For example:
\begin{enumerate}
    \item To schedule a test, a user has to share their name, email, phone number, ethnicity, country of birth, and recent travel history. Although it differs based on the testing site and country, this personally identifiable information often sits on web servers or in handwritten forms at testing campsites.
    \item Some private testing companies collect extensive user information about the user despite its irrelevance to COVID-19 testing. This can vary significantly between testing sites and countries. The use of anonymous randomized identifiers can reduce these risks.
    \item Contact tracing is the single public health intervention that has been most effective in identifying potentially infected individuals. These efforts often neglect thorough safeguards of individual privacy.
    \item COVID-19 results are often conveyed to users on websites that require no more than name and birthdate to look up the results, an obvious privacy concern. 
\end{enumerate}

The privacy breaches outlined above can form the basis for more significant invasions of patient privacy \cite{27}, allowing disingenuous parties to track patient schedules, economic status, and social contacts. The potential misuse of this information can dramatically impact patient safety and well-being. 

\subsection{Communication}
Testing and contact tracing remains crucial to the public health workforce around the world in identifying, assessing, and managing individuals to contain the spread of COVID-19. But for many people, coming forward to get tested and revealing the personal information of friends, family, and close associates remains a challenge due to privacy reasons. The COVID-19 upsurge has given rise to discriminatory behaviors towards anyone discerned to have been in contact with the virus \cite{30}. This social stigma discourages individuals from receiving COVID-19 tests, adding another layer of complexity to public health policies and data collection efforts surrounding the pandemic.

\subsection{Economy}
The impact of COVID-19 on the economy has been significant. Due to COVID-19, several small and large-scale businesses have been negatively affected. The virulent and rapid spread of COVID-19 has required many organizations to change their work and marketing strategies. Small case businesses have been most significantly affected \cite{5}. 43\% of businesses were temporarily shut down in earlier stages of the pandemic, and the employment rate dropped down by 40\% in the United States. This has raised questions about the financial stability of small businesses as many have needed to acquire additional debts, temporarily cut down or furlough their workforce, and ultimately shut down in many cases. A recent analysis \cite{6} has provided 3 potential future economic scenarios in the setting of COVID-19: global economic slowdown, global recession, or a quick reformation of the economy. The last relies upon the implementation of public policies including widespread testing, mandated face coverings, social distancing, and new sanitization standards in order to facilitate a return to normalcy for the economy. In scenarios in which transmission of the virus cannot be curbed, economic slowdowns and recessions are far more likely.

%% file: content/future_possibilities.tex
\section{Future possibilities for Testing}

Researchers are in the preliminary phases of exploring multiple novel methods, focused on improving the accuracy, speed, cost and complexity of COVID-19 tests. One such initiative based on various biological innovations is STOPCovid \cite{21}, working to offer rapid at-home tests with simpler kits. Researchers are also exploring testing methods that leverage machine learning to detect COVID-19 from signals such as cough noises, lung and respiratory performance, and vocal cord strength \cite{18}. These methods have the potential to substantially simplify the testing process, allowing for increased accessibility and mass population testing. 

%% file: content/conclusion.tex
\section{Discussion and Conclusion}
As the COVID-19 pandemic continues to rapidly grow in the United States and across the world, new testing methodologies will continually be developed. In this early draft, we present a detailed review of the clinical landscape of COVID-19 testing and establish the metrics by which future tests might be measured and understood. Test sensitivity is only one factor that must be considered in combination with a test’s cost, speed, diagnostic potential,

Using this framework, we describe the current mainline diagnostic and serological testing methodologies available as well as their relative advantages and use cases. Further, we extend our analysis to novel, unconventional, and alternative methods of testing, which could reduce the time and cost involved to obtain results. 

We also discuss the potential privacy concerns relevant to current testing approaches. Given the tradeoffs between benefits vs harms of testing, the landscape remains open to new testing strategies, especially those that minimize the risk to an individual’s privacy. New testing methods and tools, such as STOPCovid and Cosware, might be used to supplement current COVID-19 testing strategies while maintaining user privacy.